# Does it matter if you answer slowly?

*Srdjan Verbić[1]*

**Abstract**

In this paper, we have analyzed item response times measured at a large-scale unspeeded low-stakes test for primary-school students. We have demonstrated the existence of significant difference in the response time for boys and girls as well as difference in response time of correct and incorrect answers on this test. We have also demonstrated existence of the warm-up effect for this test. The results show that responses given by girls exhibit much greater warm-up effect and that difference appears to be the most important cause of the difference on the test level.

## *Introduction*

### Item response time

Computer-based testing enables us to record duration of students' responses for each test item. Item response time (RT), or item latency in some literature, is generally defined as time elapsed between presenting a question on a computer screen and response to that question.

Response time data can be valuable additional source of information on both: the characteristics of the test takers and the test. Studies in the past two decades explored the use of item response time for different applications like assessing motivation (Beck, 2004; Wise & Kong, 2005), investigating response strategies (Schnipke & Scrams, 1997), investigating test security issues (Meijer & Sotaridona, 2005; van der Linden & van Krimpen-Stoop, 2003), investigating test item cognitive characteristics (Gvozdenko & Chambers, 2007), or assessing and controlling the speededness of a test (Bridgeman & Cline, 2000; van der Linden, Scrams, & Schnipke, 1999). Most of the studies concerning item response time are dealing with high-stakes tests, which are speeded by setting the time limit for the entire test. The research of response time in unspeeded low-stakes tests is less common. In this kind of tests, we can expect somewhat different responding behavior of test takers like lower motivation, which is an issue that the researchers are most concerned about (Lee & Chen, 2011), or relaxed tempo of responding. This study is focused only on those behaviors that are relevant for low-stakes unspeeded testing conditions, i.e. situation when students can choose their own pace for answering to the test questions.

Students' response times to the same item in a test can be different by more than two orders of magnitude. Typical RT to different items in the test do not vary that much.

---
[1] Institute for Education Quality and Evaluation, Belgrade

Some demanding open-ended items may require quite long RT from all students, while some easy multiple-choice items can be answered much faster. For the purpose of RT analysis, sometimes it is more convenient to compare RT in absolute units (e.g. if item properties are explored), while for the exploration of students' item-responding characteristic and behavior, comparison of relative RT scaled for each test item would be more convenient.

## Normalized response time

Test takers' response time for a single item has distribution that is very similar to lognormal. Many previous studies demonstrated that item response times, for a fixed item, fit to lognormal distribution better than to the other common distributions (Schnipke & Scrams, 1999; Thissen, 1983; Thompson, Yang, & Chauvin, 2009; van der Linden, 2006). Therefore, we can easily normalize response time data with natural logarithmic transformation and, consequently, create a more normal distribution required by most of statistical procedures. The logarithms of test takers' response times (log RT) should follow normal distribution for each particular item. If we denote response time of the test taker $l$ to the item $k$ with $t_{kl}$, the parameters of lognormal density functions can be estimated by taking the mean log RT ($\mu_k = \text{mean}(\log t_{kl})$) and standard deviation of log RT ($\sigma_k = \text{sd}(\log t_{kl})$) across all test takers ($l=1...n$) for each item $k=1...m$, when $n$ and $m$ represent the number of test takers and the number of items respectively. Hence, we can define normalized item response time ($\tau_{kl}$) as:

$$\tau_{kl} = \frac{1}{\sigma_k}(\log t_{kl} - \mu_k). \tag{0.1}$$

We can think of $\tau$ as of relative response time. For example, student $l$ whose response time to an item $k$ is average among the other students in the group has $\tau_{kl} \approx 0$. Students who have normalized item response time greater than zero ($\tau_{kl} > 0$) gave their responses to item $k$ slower than average and vice versa. Assuming that RT follows lognormal distributions for each item, we can define characteristic item response times ($t_k$), which corresponds to:

$$t_k = e^{\mu_k}. \tag{0.2}$$

Since $e^{\text{mean}(\log RT)} \approx \text{median}(RT)$ for distributions close to lognormal, we can think of characteristic item response times ($t_k$) also as of median values of RT for the item $k$.

## Difference between response time of correct and incorrect answers

Previous research concerning item response time in various testing situations demonstrated that response time of correct answer is generally shorter than RT of incorrect (Bergstrom, Gershon, & Lunz, 1994; Chang, Plake, & Ferdous, 2005; Chang, Plake, Kramer, & Lien, 2011; Hornke, 2000, 2005; Thompson et al., 2009; Troche & Rammsayer, 2005). It is also noticed that RT, on average, increases with the item difficulty (Bergstrom et al., 1994; Bridgeman & Cline, 2000; Wheadon & He, 2006). To our knowledge, it is not well established whether and how RT of correct and incorrect answer, or their difference, varies with item difficulty. Such a behavior should depend greatly on the stakes of the test. For students taking low-stakes test, it is critical how much effort they would put in answering difficult or lengthy questions. It is more likely that they would give up solving the problem and guess if they do not recognize the correct answer after reasonably long time (Schnipke & Scrams, 1997). Guessing behavior should be more emphasized for difficult items, which is expected to affect difference between response time of those who know the answer and those who do not.

**Gender difference**

Previous studies concerning differential response time for different subgroups were focused mainly on high-stakes tests and speeded conditions (Llabre & Froman, 1987; Schnipke & Pashley, 1997). Wise and his colleagues (2004) explored rapid-guessing behavior at low-stakes tests where examinees responded quickly because of the lack of motivation to work hard on some items. They reported that girls at low-stakes test do not respond rapidly as much as boys do. In more relaxed situations where students do not really rush to finish the test, it is also reported that boys respond quicker than girls do (Verbić & Tomić, 2009). Schnipke (1995) had studied gender differences relative to the effort; she found that rapid-guessing behavior was more common among male examinees on an analytical test. In addition, rapid guessing was more common among female examinees on a quantitative test, and equally common on a verbal test.

Generally, various subgroups, not based only on gender but also on ethnicity or mother tongue, may differ in the dynamics at which they respond to test items. This difference in speed could be the cause of bias of item parameters and estimation of test reliability, as well as the cause of Differential Item Functioning (Oshima, 1994).

**Warm-up effect**

Previous studies had also confirmed that position of an item in a test affects the response times of the test takers (Halkitis & Jones, 1996; Swanson, Case, Ripkey, Clauser, & Holtman, 2001; van der Linden, Breithaupt, Chuah, & Zhang, 2007). In all these studies students tended to respond slower at the beginning and faster toward the end of the tests. The change in the response time due to position appears to be small and do not have any systematic impact on the item scores. We can say that the test takers in the beginning of the test feel inclined to spend more time on the items than they actually needed (van der Linden et al., 2007). Such a pattern is named by Bergstrom et al. (1994) the warm-up effect.

**Research questions**

Using ideas and results of previous research in item response time, we have tried to evaluate and answer the following questions for an unspeeded low-stakes test:
      1 Is there a difference between response time of correct and incorrect answers?
      2 Does item response time depend on item difficulty or item's position in a test?
      3 Is there a difference between response time for boys and girls?

*Method*

**Instrument and sample**

Methods and data provided in this paper are results of secondary analysis of a low-stakes computer-based test for elementary school students. Main results of the test were previously reported in (Verbić, Tomić, & Kartal, 2009). Test consisted of 32 multiple-choice and multiple-response items. Questions for the test were selected from the annual test in subject Nature & Society for fourth-graders. Purpose of the test was to enable pupils to participate in a national-level study using their school computers and test their own knowledge. Fifty schools applied for the testing and 926 pupils from these schools participated in. The study was realized in unspeeded conditions where all students finished their tests in less than one hour.

**Test design**

Items labeled with #1, #2, …, #32 were grouped in four sections (A, B, C, and D) consisting of eight items each. These sections were used to construct four variants of the same test with the rolling-sections order displayed in Table 1. For example, the pupil who received test variant number 2 has started the test with section B and ended with section A.

**Table 1: Rolling-sections design of four variants of the test**

|           | $1^{st}$ period | $2^{nd}$ period | $3^{rd}$ period | $4^{th}$ period |
|-----------|-----------------|-----------------|-----------------|-----------------|
| Variant 1 | A               | B               | C               | D               |
| Variant 2 | B               | C               | D               | A               |
| Variant 3 | C               | D               | A               | B               |
| Variant 4 | D               | A               | B               | C               |

All four variants were evenly distributed among test takers. Such a design enables us to examine items functioning and the duration of test taker's responses when the same item for different students appears in different phases or periods of testing. That way we can diminish the effects of item ordering on the item response time for a particular item.

**Measuring response time**

The computer-based testing was developed and administered using Moodle course management system (Dougiamas, 2001) with an additional module written for capturing response times. Each question was presented on a separate screen, which allowed measuring of time that examinees spent on a particular question. Test takers could also omit items (i.e., see an item without answering it and proceed to another), and they could go back latter to previously viewed items and change their answers. The test was administered in this way in order to make it as similar as possible to the paper-and-pencil mode of test delivery. The recorded response time for an item was the total time spent on the item during all attempts as it was proposed by Schnipke & Scrams (1997). For the purpose of this study, the time of the first approach and the accumulated time for all subsequent approaches to a question were recorded for each test taker and each item. The response times were acquired with one-second accuracy.

**Item response theory and determination of item parameters**

Item Response Theory (IRT) is more useful for the analysis of large-scale low-stakes tests than Classical Test Theory because of IRT ability to deal with missing data. At low-stakes tests, many students omit items because of the lack of motivation to work hard on all the items. Since we cannot say whether students who omitted several items are less able or not, we should use the advantage of IRT to estimate student's ability based on answered items only. Because of item parameter estimation stability, we have used two-parameter (2PL) IRT model for estimating item parameters and expected a posteriori estimation algorithm proposed by Bock and Mislevy (1982) for the estimation of test takers' ability.

## *Results*

Examinees' response times (RT) to any particular item in the PD09 test can differ more than two orders of magnitude. Distributions of RT to all items in the test are presented on Figure 1. Median value of RT for the item where examinees responded fastest (#28) is 21 second, while for the item where examinees responded slowest (#25) it is nearly four times greater (80 seconds). The variance of RT within an item is obviously much greater than the variance between items.

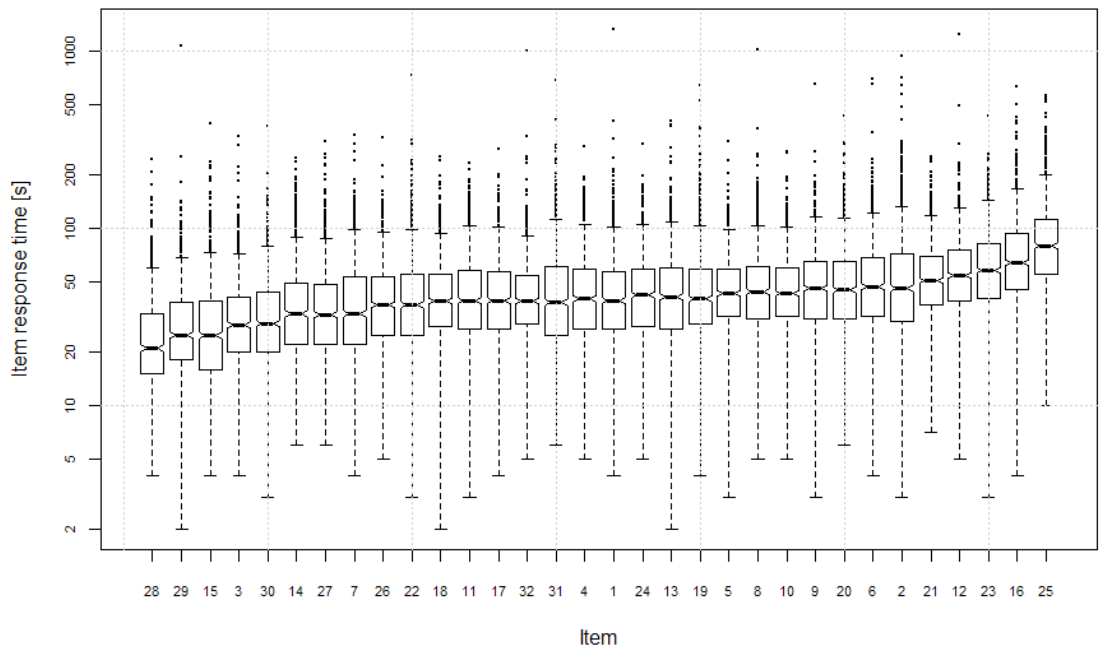

Figure 1: Distributions of item response times for all items in the PD09 test

Simple analysis of response time (RT) data obtained from the PD09 test shows that average item RT of correct answers (median value 38 s) was much lower than RT of wrong answers (median value 45 s). Using *t*-test for log RT, we can quantify difference between log RT of correct answers and wrong answers, on the item level, and clearly demonstrate statistical significance of the difference: $t(21833)=-24.3$, $p<2e-16$. Curiously, we have found that log RT practically does not correlate with students' estimated ability level: $r<0.02$.

We have found also that difference in log RT between correct and wrong answers depends much on the item difficulty. This relationship is presented on Figure 2. In this test, we have observed that 17 out of 32 items had significantly lower log RT of correct than of wrong answers. These items are represented by sign ▼ on Figure 2. All these items were easy (IRT difficulty parameter $b<0$). We also had three items were correct answers had significantly higher log RT than responses of wrong answers (represented by sign ▲). All these items were difficult (*b*>0). Correlation between the difference in log RT and item difficulty is *r*=0.73 (*p*<3e-6).

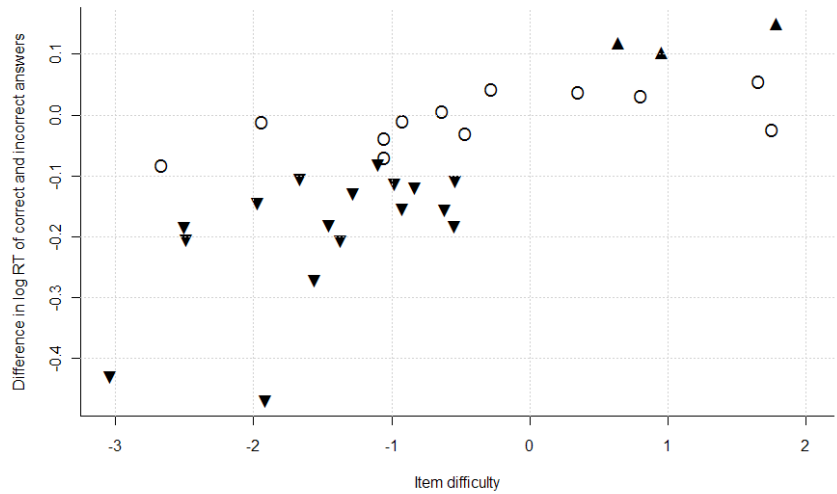

**Figure 2: Difference between log RT of correct and incorrect answers to an item depending on the item difficulty (*b*): *r*=0.73, *p*<3e-6. Upper (▲) and lower (▼) triangles represent items where log RT of correct answers was significantly lower or higher than of wrong answers. Hollow circles represent items where differences were not statistically significant.**

Response times captured on this test show also that boys, on average, respond significantly faster than girls do. Median time needed by boys to finish the entire test was 1525 seconds while the girls, on average, finished 81 second later. If we compare all student-item encounters, we see that difference between log RT for boys and girls is statistically significant: $t(28753)=-6.6$, $p<4e-11$. The logarithm of item RT was considerably higher for boys then for girls at 11 out of 32 items. There were no items where girls' log RT was significantly higher. In Table 2 are given characteristic response times of correct and incorrect answers for both boys and girls. Characteristic response time for each group is calculated as $e^{mean(log\ RT)}$.

**Table 2: Characteristic response time of correct and incorrect answers for boys and girls given in seconds. Estimated standard errors are given in parentheses.**

|  | Characteristic response time [s] | |
|---|---|---|
|  | Correct answers | Incorrect answers |
| Boys | 36.8(2) | 44.7(2) |
| Girls | 38.8(2) | 46.3(2) |

Shifting items' positions in the test enabled us to compare response times for an item at different positions. Response time to an item is much longer when the item is being displayed as the first item than when it is positioned elsewhere in the test. Items #13, #27, #15, and #16 have been positioned at each 1st, 9th, 17th, and 25th place in the test depending on the test variant. Response times to all these items were much longer when students encounter with them at the very beginning of test (Table 3). For item #27 we can see that median response time is twice longer when it is at 1st position than at 9th, 17th, or 25th position. Similar occurrence, but smaller in magnitude, could be seen for 2nd and 3rd position in the test also.

Table 3: Median response time to four items appearing at 1st, 9th, 17th, or 25th position in the test for 4 test variants A, B, C, and D

|  | Median item response time [s] | | | |
| --- | --- | --- | --- | --- |
|  | 1st position | 9th position | 17th position | 25th position |
| Item #13 | 69 | 36 | 34 | 34 |
| Item #27 | 56 | 28 | 28 | 28 |
| Item #15 | 30 | 25.5 | 24 | 23 |
| Item #16 | 85 | 60 | 58.5 | 58 |

In order to make comparison between response times for items at different positions, we had to normalize response times for all the items. Therefore we calculated mean value of normalized item response times ($\bar{\tau}_i$) for all four items appearing at certain position *i* in the test. These results, separately for boys and girls, are presented on Figure 3. If there were no effect of item position to the response time, values of $\bar{\tau}_i$ for all positions would be equal to zero within the error margin. However, we can see that $\bar{\tau}_i$ has values above zero for the first seven positions. Mean normalized item response time has the highest value in the very beginning of the test, and then decreases quickly. The value of $\bar{\tau}_i$ increases again in the end of the test. Accelerating in the beginning and decelerating in the end are labeled here as *warm-up* and *cool-down* effects.

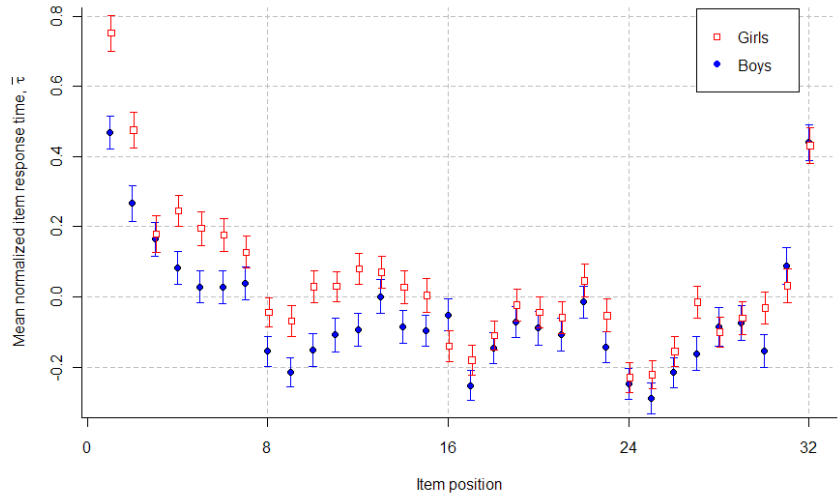

Figure 3: The mean normalized item response time ($\bar{\tau}$) depending on the item possition in the test. Error bars represents estimated standard errors. We can see that $\bar{\tau}$ is much greater in the beginning and the very end of the test.

The effect of item position on RT is not exactly the same for boys and girls. In the beginning, girls spend more time to answer questions than the boys do. Later, that difference decreases. In the first period, girls, on average, spend 41 second more to answer questions than boys do. This difference is statistically very significant: $t(893)=-4.4$, $p<2e-5$. In the following three periods, girls are also slower than boys, but that difference is not significant. In order to compare RT for each position, we compared mean normalized item response times instead of mean response times. From Figure 3 we can see that mean normalized RT for girls is significantly higher than for boys in the beginning of the test. This difference appears to decrease gradually with item position. Correlation between difference in $\bar{\tau}_i$ for boys and girls and the item position $i$ is $r$=-0.62, $p<0.0002$.

## *Discussion*

Distribution of an item response time depends on the item difficulty and its position in the test. However, these characteristics can explain only a small part of item RT variance. Therefore, knowing an item difficulty and its position in a test cannot help us much to predict response time of a particular examinee to that particular item. Rather, it can help us to recognize some general RT patterns and estimate test characteristics regarding responding dynamics and duration of the test.

Each item has its own characteristic response time, but all their distributions fit well to the lognormal function. Utilizing this property of RT we can normalize item response times for all items. That enables us to represent RT for all items on a common scale and explore more subtle effects influencing item response time.

Analysis of normalized item RT shows that girls, on average, respond slower than boys do to great majority of items in the PD09 test. For one third of items, this difference is statistically significant. The existence of four variants of the same test enables us to analyze typical response times for each of four periods of testing. Mean normalized item RT has the highest value in the very beginning of the test, and then decreases quickly. The value of mean normalized RT increases again at the end of the test. This means that students start slow and accelerate quickly until reach "working speed". Finally, when they approach to the end, they slow down and take their time to answer carefully the last few questions. This behavior demonstrates existence of both the warm-up and cold-down effect in unspeeded testing conditions provided for the PD09 test.

There is no evidence that difference in response time causes differential item functioning on the test PD09. Difference in response time for boys and girls did not affect their achievement. Therefore, we cannot tell that boys perform better than girls or vice versa, but we certainly can say that boys, on average, respond faster.

## *Conclusions*

Item response time is outcome of complex and yet unknown internal processes in student-item interaction. Looking from the outside, these outcomes look like random values obtained from lognormal distributions characteristic for each particular item. Information about response time, probably, cannot help us much to estimate examinee's ability with greater precision directly, but could help us to write better items and design more efficient and equitable tests. Primary intention of this paper was to make a sound contribution to the understanding of students-item interaction and the response dynamics.

The evidences that correct answers take less time than incorrect are often reported. Here is given an example where such a difference clearly depends on the item difficulty. The reason for this occurrence might be students' motivation to put effort in solving hard problems on a low-stakes test. If they do not find solution in reasonable amount of time, they give up and guess. That is why the most of incorrect answers to hard questions have short response times. Curiously, this test takers' behavior has nothing to do with their latent ability. Further research in item RT might help us to observe students, among both: high- and low-achievers, who give unreliable answers and predict effects of this behavior on the test properties.

Difference in response times of boys and girls implies that test can be unfair if examiners impose a time limit on a short test. This difference seems to be a consequence of differential behavior of boys and girls in the beginning of a test. This paper shows an example of unspeeded test where such a difference does not induce difference in the achievement of these two groups of examinees. Better description of conditions where response time does not influence achievement requires further research. Additional results on differential item response time consequentially could improve equity of assessment.